# Omnidirectional Ventilated Acoustic Barrier

Hai-long Zhang[1,2], Yi-fan Zhu[1,2], Bin Liang[1,2*], Jing Yang[1,2], Jun Yang[3], and Jian-chun Cheng[1,2]

[1]*Key Laboratory of Modern Acoustics, MOE, Institute of Acoustics, Department of Physics, Nanjing University, Nanjing 210093, P. R. China.*

[2]*Collaborative Innovation Center of Advanced Microstructures, Nanjing University, Nanjing 210093, P. R. China.*

[3]*Key Laboratory of Noise and Vibration Research, Institute of Acoustics, Chinese Academy of Sciences, Beijing 100190, P. R. China.*

[*]Correspondence and requests for materials should be addressed to B. L. (email: liangbin@nju.edu.cn)

**Abstract**

As an important problem in acoustics, sound insulation finds applications in a great variety of situations. In the existing schemes, however, there has always been a trade-off between the thinness of sound-insulating devices and their ventilating capabilities, limiting their potentials in the control of low-frequency sound in high ventilation environments. Here we design and experimentally implement an omnidirectional acoustic barrier with planar profile, subwavelength thickness ($0.18\lambda$) yet high ventilation. The proposed mechanism is based on the interference between the resonant scattering of discrete states and the background scattering of continuous states that induces Fano-like asymmetric transmission profile. Benefitting from the binary-structured design of coiled unit and hollow pipe, it maximally simplifies the design and fabrication while ensuring the ventilation for all the non-resonant units with open tubes. The simulated and measured results agree well, showing the effectiveness of our proposed mechanism to block low frequency sound coming from various directions while allowing 63% of the air flow to pass. We anticipate our design to open routes to design sound insulators and to enable applications in traditionally unattainable cases such as those calling for noise reduction and cooling simultaneously.

As an important problem in acoustics, sound insulation [1-9] finds wide applications in diverse scenarios ranging from noise control to architectural acoustics. In the existing schemes, however, there has always been a trade-off between the thickness of sound-insulating devices and their ventilating capabilities, limiting their potentials in the control of low frequency sound in high ventilation environments. Conventionally, sound insulation can be realized by both active [8-9] and passive methods [3-7]. Compared with active methods that need complicated and costly electronic systems, the use of passive structures provide simple solutions much easier to apply in practice. Yet the passive methods generally have to rely on impedance mismatch by insertion of layered materials, which would be bulky in terms of wavelength if realized with natural materials [1]. Although the advance of metamaterials [10-21] has overcome the problem of limited acoustical properties available in nature and enabled substantial reduction in both the thickness and mass density of sound-proof structures such as by using membrane-type metamaterials [3-6], there is still a fundamental limit that the inserted natural or artificial materials necessarily lead to discontinuity of the surrounding air, making them not practical in environments in need of ventilation. Despite the recent emergence of open structures for sound insulation, they need to decorate the inner boundaries of a waveguide with metasurfaces [14-16, 19-21] for generating anomalous reflection and therefore have to be bulky-sized, angular-dependent and inapplicable to free space [7]. To date, mechanism for effectively blocking omnidirectional low frequency sound while keeping high-efficiency ventilation property is still to be explored as a result of its significance to the design and application of sound insulators.

In this Letter, we propose to design an acoustic barrier that simultaneously enables high-efficiency transmission of other entities (viz., light, fluid, etc.) and blocks omnidirectional low frequency sound, as schematically illustrated in Fig. 1. The mechanism is that, instead of relying on impedance mismatch by inserting other materials that necessarily lead to discontinuity of surrounding air or production of anomalous reflection with metasurfaces that need to be decorated within a waveguide much longer than the wavelength, we introduce the resonant scattering of discrete states with the coupling of the background scattering of continuous states. Through interaction of the two modes, transmission dip will occur in at the Fano-type resonance [22-24]. The resonant unit is fabricated with labyrinthine structure that substantially downsizes the thickness of the proposed device for insulating low frequency sound. In addition, the non-resonant unit constructed by hollow pipes contributes the continuous sound field and flow field. As illustrated in Fig. 1, the metamaterial-based barrier is composed of two different unit cells whose structural outline and parameters are exhibited.

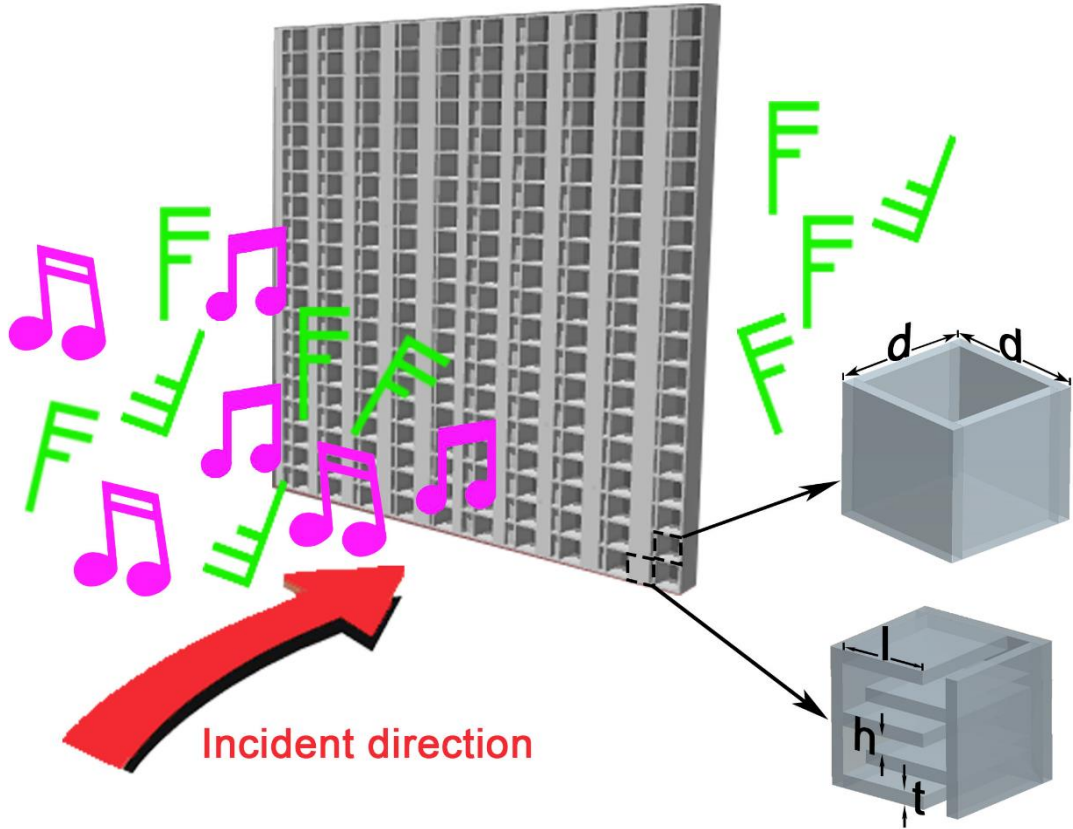

FIG.1 Schematic of the proposed acoustic barrier with planar profile and subwavelength thickness that insulates the propagation of sound while allowing the transmission of fluid. The wind signal represents the wind and the 'note' represents the acoustic wave in both sides of the planer surface. The two units which can be implemented by using a labyrinthine-type metastructure and a straight hollow pipe respectively are enlarged.

It has been extensively proven that the type of metastructure with labyrinthine-like geometry serves as good candidate for acoustic metasurface and resonator due to the capability of causing full $2\pi$ delay in the propagating phase within a distance much smaller than the wavelength when the sound passes through it, for which the phase delay can be freely tuned via adjustment of its structural parameters. Here these parameters are chosen as $d=10$mm, $l=5.6$mm, $h=1.25$mm and $t=1$mm. As shown in Fig. 2(a), the coiled structure has a transmission peak and phase abruption at the frequency of 5600Hz where the Fabry-Perot resonance occurs [14, 15]. On the other hand, the hollow pipe unit always allows a unity transmission as long as this unit has a subwavelength size. In our proposed device formed by combining these two kinds of components together, therefore, the interference between the resonant scattering of discrete states and the background scattering of continuous states, which can be clearly observed in Fig. 2(c), results in Fano resonance as evidence by changing from the symmetric Lorentzian profile [22] to asymmetric shape of transmission curve [22-24] which now has a transmission dip accompanied by a unity transmission in a frequency below as shown in Fig. 2(b). Note also that the location of the transmission peak is almost at the resonance frequency of the coiled structure (albeit with a slight red shift by about 4% of the resonance frequency, i.e., from 5600 Hz to 5400 Hz in the current case, which agrees quite well with the value reported in the literatures [25-26]. In addition, as the transverse dimensions of the units are subwavelength, such asymmetric peak-dip behavior is not sensitive to the incident angle of sounds. Red arrows in Fig. 2(c) show the direction of the sound intensity at the two different frequencies. Length of the arrows are in the logarithmic form to show the intensity of sound wave. At the dip's frequency, 5900Hz, phase of the labyrinthine unit is obviously opposite from the hollow pipe which can also be seen from the sound pressure field in Fig. 2(d). It is the interaction between the resonant scattering of discrete states and the background scattering of continuous states that weakens the energy radiated to the far field.

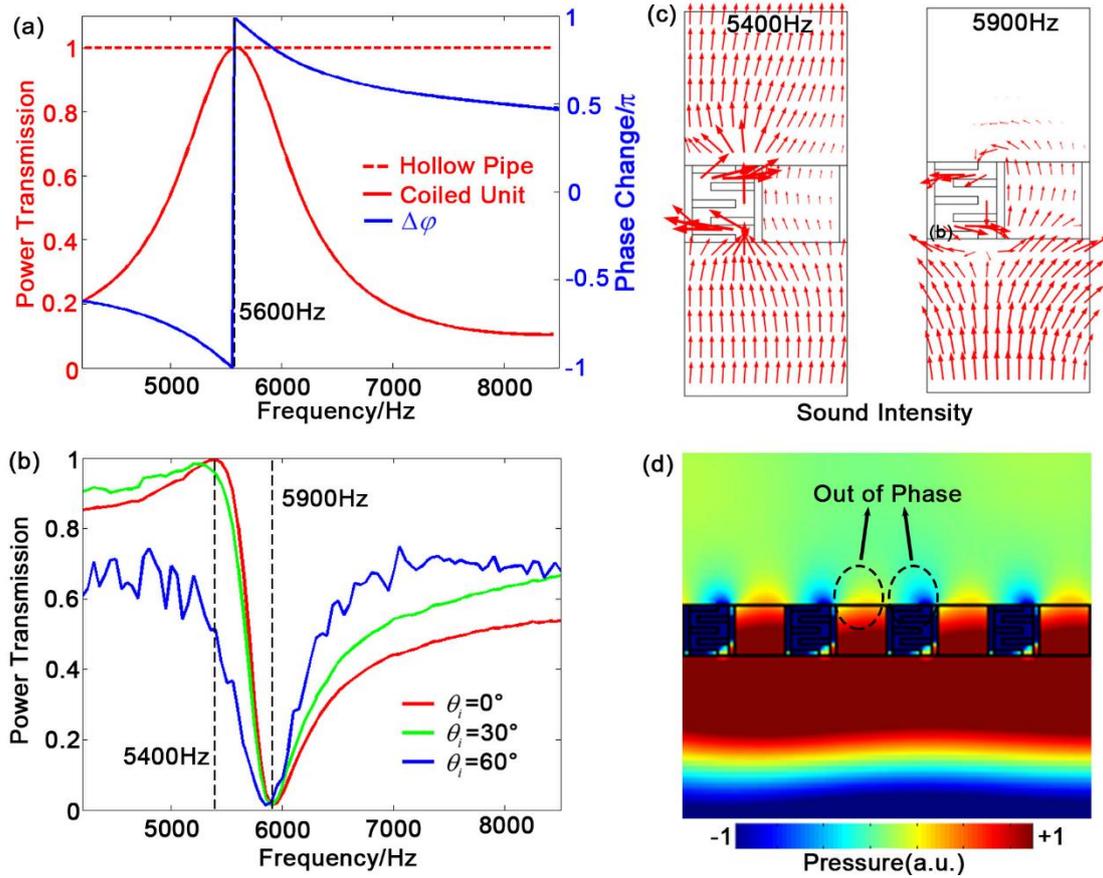

FIG.2 (a) Transmission spectra of the coiled structure, hollow pipe and phase delay of the resonant unit versus frequency. (b) Frequency dependence of transmission for three particular incident angles of $\theta_i=0°$, 30° and 60°, respectively. (c) Sound intensity distribution in an individual period at two particular frequencies of 5400Hz and 5900Hz where the transmission peak and dip occurs, respectively. (d) Acoustic pressure distribution in the near field shows the surface-bound mode which originates from the destructive interference of the two transmitted fields.

Considering the random incident angles the acoustic signals may have in practical applications, it is significant to inspect the angular dependence of the transmission property of our proposed acoustic barrier. We simulate the transmission spectra for both the conditions of plane wave incidence and point source radiation [27]. In the numerical simulations, the solid material used for constructing the metastructure is chosen as Acrylonitrile-Butadiene-Styrene (ABS) plastic for which the sound speed and mass density are $c = 2700$m/s and $\rho = 1180$kg/m$^3$ respectively, which will be employed for the sample fabrication in the experiment. The simulated acoustic field shows a dramatic reduction in the transmission at 5900Hz regardless of the incident angle of plane wave, demonstrating the near-omnidirectional functionality of our acoustic barrier, which can be ascribed to result from the interaction of the two resonant and non-resonant modes. Eventually, such coupling of resonance between neighboring unit cells causes the total reflection of incident sound energy on the surface of the acoustic barrier and thereby leads the transmitted wave to approach zero, which is evidence by the simulated sound field distributions shown in Figs. 3 that reveal the interference pattern in the incident side and the vanishing transmitted wave field on the opposite side when a plane wave impinges on the designed acoustic barrier from different angles. In Figs. 3(a)-3(c) the frequency is chosen as 5900 Hz and the structural parameters are the same as Fig. 2(b). For a precise estimation of the strength of transmitted wave, we use a square cross-section waveguide with rigid boundary in the transmitted side and obtain the transmission coefficient by integrating the normal component of transmitted energy flux over the whole cross section which is also the scheme for measuring the transmission coefficient to be employed in the following experiment. In the output side of the barrier, the two opposite boundaries along the *y* direction are set as hard boundaries, in addition to this, all the other boundaries are radiation boundaries. The total reflection can be observed in the input side of the barrier for the almost zero energy transmission. A standing wave field in the incident space takes shape for the constructive interference between the incident wave and reflected wave. Furthermore, the simulated pressure distribution for an acoustic barrier impinged by a point source and the sound pressure level in the far field as a function of incident angle at different frequencies also show the omnidirectional shielding effect of the acoustic barrier [27]. The transmission loss is about 25dB between the total transmission and reflection situations.

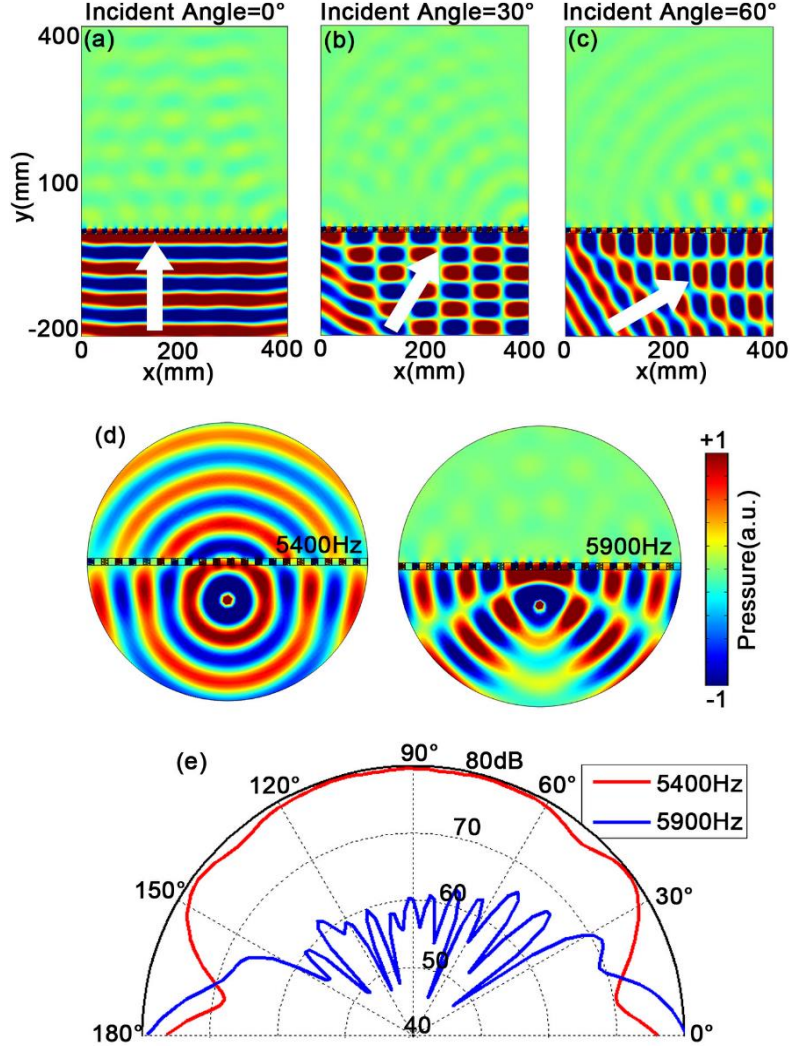

FIG. 3 The simulated acoustic pressure distributions caused by the proposed structure when illuminated by a plane wave with incident angles: (a) $\theta_i=0°$ (b) $\theta_i=30°$ (c) $\theta_i=60°$. (d) Spatial distribution of acoustic pressure when the proposed acoustic barrier is impinged by a point source driven by two particular frequencies of 5400Hz and 5900Hz where the transmission peak and dip occurs respectively. (e) Angular dependence of the sound pressure level in the far field for acoustic barrier at different frequencies.

Experiments to verify and demonstrate our metasmaterial based sound insulation are conducted in the anechoic chamber with samples fabricated via 3-D printing technique (Stratasys Dimension Elite, 0.177 mm in precision). The experimental setup and the samples are shown in Figs. 4(a), 4(b) and 4(c). A loud speaker is placed in front of the sample located on a rotating stage. The distance between the loudspeaker and the sample is set as 60cm such that it is sufficiently to treat the loudspeaker as a sound source emitting a plane wave within the frequency range of interest (i.e., 4200 – 8500Hz). The transmitted acoustic intensity is measured by using a 1/4-inch-diameter Brüel & Kjær type-4961 microphone.

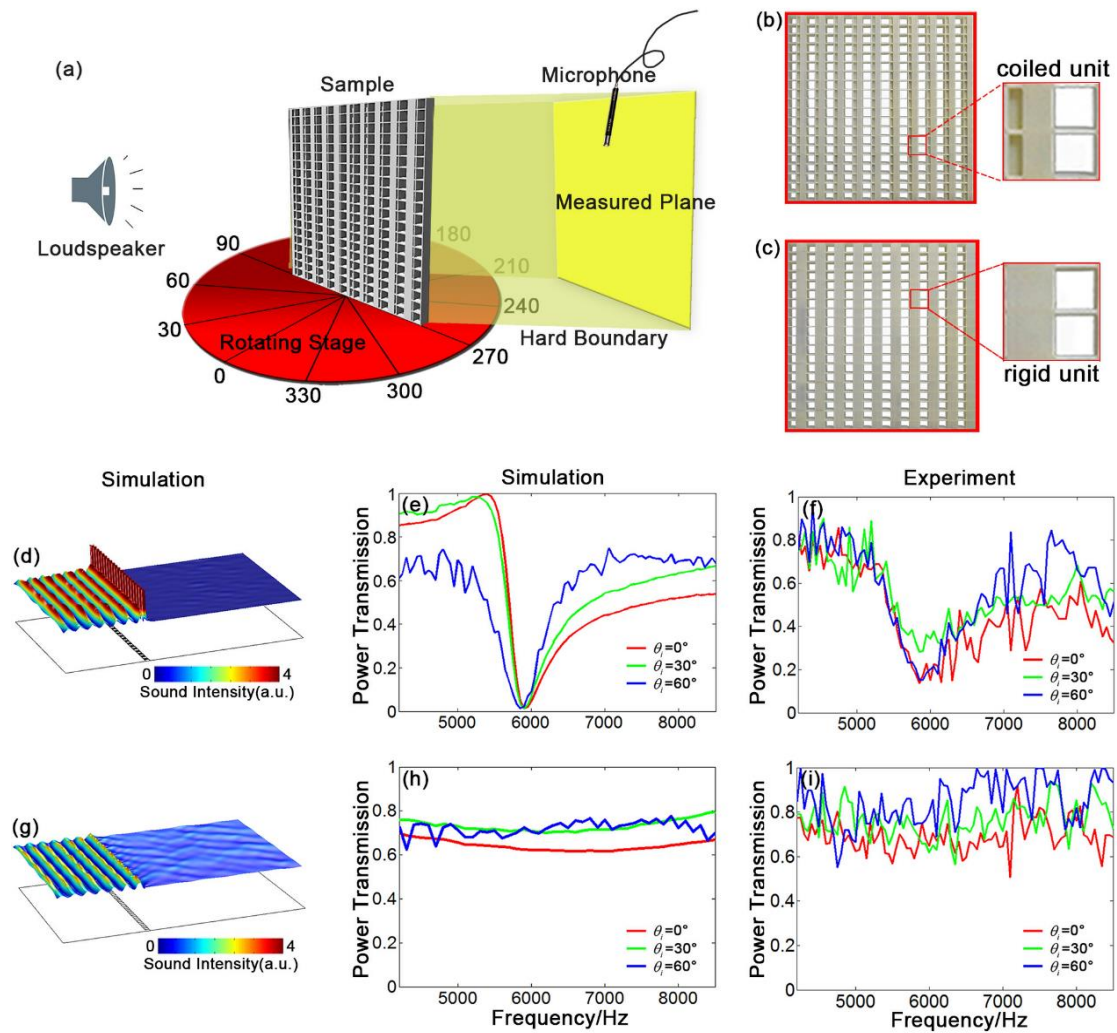

FIG. 4 (a) Schematic of the experimental setup. Photograph of the acoustic barrier sample (b) and the referenced sample with rigid unit cells (c). Simulated spatial distribution of acoustic pressure amplitude at the frequency of 5900Hz for (d) the designed acoustic barrier and (g) a reference with rigid units. Simulated (e, h) and measured (f, i) frequency dependences of power transmission for coiled metastructure (e, f) and a reference barrier with rigid units (h, i) for three particular incident angles: $\theta_i$=0°, 30° and 60°.

Figures 4(d) and 4(g) display the comparison between the spatial distribution of sound pressure amplitudes at normal incidence and frequency of 5900Hz for the designed coiled acoustic barrier and a referenced sample of rigid units. The sound intensity distributions clearly exhibit a huge amplification at the surface of coiled structure due to the coupling between neighboring unit cells, leading to the total reflection of incident energy and a nearly complete diminish of the transmitted waves. Good agreements are observed between the simulated and measured results of transmissions as functions of frquencies for different incident angles, demonstrating that the propagation of incident wave is virtually blocked nearly omni-directionally despite its subwavelength thickness and holey structure. It is worth pointing out that we have considered the viscous effect in the simulations and the slight discrepancy between the numerical and measured results should come from the imperfect sample fabrication and non-zero reflection at the end. For a quantitative estimation of the ventilation property of the resulting device, we have also measured the air flow rate of the sample in addition to its acoustical property using the same measurement setup. To measure the output air flow in the experiment, we place a wind speed meter (type TECMAN TM856) at the exit of the rigid tube and make it parallel to the sample with a distance of 30cm in between to avoid the diffusion of air that influences the measurement of flow velocity. An electric fan (type KONKA 25HY38) is placed in front of the ventilated acoustic barrier to generate steady airflow with the driving voltage being 220V. The measured data show that the measured air flow rate without and with the acoustic barrier are 2.72m/s and 1.72m/s, respectively, demonstrating a high ventilation effect of our design that allows 63% of the airflow to pass through the metamaterial.

In summary, we report the simulated and experimental verification of an acoustic barrier with subwavelength thickness ($0.18\lambda$) capable of blocking low frequency sound with random incident angles (ranging from normal incidence to grazing incidence) while maintaining high ventilation ability (63% ventilation rate). Such extraordinary feature comes from the interference between the resonant scattering of discrete states and the background scattering of continuous states, that is, the Fano-resonance. At the frequency of Fano-resonance, phase of the two units are clearly opposite that weakens the transmitted energy. Both the numerical simulations and experiments demonstrate the effectiveness of the subwavelength ventliated acoustic barrier. In this condition, the realization of omnidirectional shielding of acoustic wave in such a compact and opened manner adds capabilities for manipulating acoustic waves without impeding airflow. In the near furtute, the approach for sound insulation and the sample we proposed here should help solve the problem in the area of environmental noise control and architectural acoustics, e.g. acoustic barriers on the highways and central processing units with the need of noise reduction and cooling simultaneously.

This work was supported by the National Natural Science Foundation of China (Grant Nos. 11634006 and 81127901) and A Project Funded by the Priority Academic Program Development of Jiangsu Higher Education Institutions.